\documentclass{aa} 
\usepackage{amsmath}
\usepackage{longtable}

\usepackage{graphicx}
\usepackage{txfonts}
\usepackage{orcidlink}
\usepackage{hyperref}
\usepackage{url}
\usepackage{comment}
\usepackage{multicol}
\usepackage{xcolor}
\begin{document}

   \title{Analysing the flux stability of stellar calibrator candidates with TESS} 

   \author{E. Tonucci
          \inst{1,2,3}\,\orcidlink{0009-0006-4370-822X}
          \and
          T.A. van Kempen\inst{2}\,\orcidlink{0000-0001-8178-5054}
          \and
          J.-P. Beaulieu\inst{4}\,\orcidlink{0000-0003-0014-3354}
          \and
          L. Bernard \inst{4}
          }

   \institute{Leiden Observatory, Leiden University, P.O. Box 9513, 2300 RA, Leiden (Netherlands)\\
         \email{tonucci@strw.leidenuniv.nl}
         \and
         Delft University of Technology, Faculty of Aerospace Engineering, Kluyverweg 1, 2629 HS, Delft (Netherlands)
        \and
             SRON Netherlands Institute for Space Research, Niels Bohrweg 4, 2333 CA, Leiden (Netherlands) \\
             \email{t.a.van.kempen@sron.nl}
         \and 
        Institut d’Astrophysique de Paris, CNRS, Université Pierre et Marie Curie, Boulevard Arago 98B, 75014, Paris (France)
             }

   \date{Received: 5 November 2024 / Accepted: 8 February 2025}

 
  \abstract 
   {The ESA space mission Ariel requires bright sources that are stable at the level of 100ppm over 6 hours in order to accurately measure exoplanet atmospheres through transmission spectroscopy. To ensure this, in-flight instrument calibration can be performed by observing stellar calibrators.}
   {In this study, a stellar calibrator candidate list distributed over the sky is created and a flux variability analysis is performed to identify the best stellar calibrators for transit spectroscopy of exoplanet atmospheres with Ariel.}
   {A starting candidate sample of 1,937 solar-type stars is created using the all-sky surveys Two Micron All Sky Survey and Gaia. Using stellar light curves from the Transit Exoplanet Survey Satellite (TESS), the flux variability of each star is characterised by computing its Lomb-Scargle periodogram and reduced chi-squared. This enables the elimination of stars with detectable variability from the sample.}
   {$\sim 22.2\%$ of stars from the starting sample pass the selection as potential calibrators. These do not all necessarily meet Ariel’s stability requirement, although some will. No correlation between flux stability and stellar properties is found, as long as the correct value ranges for the parameters are chosen, e.g. a surface temperature between 5,000 and 6,300K. The only exception is stellar magnitude: Noise in TESS data increases as stars get dimmer, so, a high percentage of faint stars passes the selection since their variability is more likely hidden within the inherent TESS noise. Contrarily, stars brighter than 5mag cannot be used as calibrators.}
   {A list of 430 promising bright calibration targets distributed over the sky has been selected. These can potentially be used as stellar calibrators for the Ariel mission. Targets from this list will have to be further studied to determine which ones possess a flux stability better than 100ppm over 6 hours.}


   \maketitle
%

\nolinenumbers

\section{Introduction}\label{ch1}
Since the discovery of the first transiting exoplanet \citep{Charbonneau00}, the transit method has been very successful in finding new exoplanets (see e.g. the NASA Exoplanet Archive\footnote{\url{https://exoplanetarchive.ipac.caltech.edu/index.html}}). This method relies on the dimming of the observed flux when an exoplanet transits in front of its host star (i.e. the primary transit or eclipse) or is eclipsed (i.e. the secondary eclipse or occultation). In both cases, the change in flux can be detected if the sensitivity is large enough. In addition, this method can be used as a probe into an exoplanet's atmospheric composition, thermal structure, cloud formation, and other parameters by studying the absorption features of the differential and transmission spectra, as well as its phase curve variations \citep{Seager00, Charbonneau02, Parmentier18}.\\

Space missions have contributed significantly to the increase of our knowledge of exoplanets in the last years. Currently, JWST \citep[James Webb Space Telescope, NASA/ESA/CSA, 2021-present,][]{Gardner06} an ambitious multi-purpose space observatory, pursuits scientific objectives including determining physical and chemical properties of planetary systems. Thanks to its high-resolution spectroscopy and direct imaging instruments, it can focus on a detailed characterisation of exoplanetary atmospheres, possibly down to terrestrial planets. In the future, Ariel \citep[ESA/NASA/JAXA, to be launched in 2029,][]{Tinetti21} will perform transit spectroscopy of exoplanets in the infrared range of $0.5-7.8\mu m$, and will have a large sky coverage. It will study atmospheric chemical composition and thermal structure by targeting $\sim$1,000 exoplanets with a wide range of masses, radii, and host star types.\\

In an era of measuring exoplanet spectra via differential measurements during transit and eclipses, very strict requirements on instrumental stability are needed. This is to ensure the successful characterisation of minute flux variations and features. First, the required flux stability is part of the instrument design and then tested with ground campaigns. In addition, stability must be monitored in time during in-flight operations, to ensure the correct functioning of the instrument and that observations can be trusted. In the infrared, flux variations are between $10$ and $1,000ppm$ for secondary eclipses of Earth-sized to Jupiter-sized exoplanets \citep{Waldmann13} with a duration between 1 to 10 hours \citep{Tinetti21}. For the future mission Ariel, for example, the flux stability requirement is currently set to a maximum peak-to-peak dispersion of 100$ppm$ $(3\sigma)$ over 6 hours \citep{Petralia17}.\\
Different calibration methods can be employed. These can either be performed with sources internal to the spacecraft or by using measurements from external sources. For example, instruments on JWST employ stable internal light sources to calibrate the instrumental response during the observations in both imaging and spectroscopic modes \citep{Glasse06, Wright15, Jakobsen22}. In addition, JWST relies on regular observations of standard stars \citep{Gordon22} for flux calibration. Standard stars, also called stellar calibrators, are stars whose spectral energy distribution at the desired wavelength range is well-studied and constant in time, thus able to act as external references for flux calibration. They are used to constrain the absolute flux calibration, but also to assess instrument stability on timescales of exoplanetary transits, particularly relevant for transit spectroscopy \citep{Tinetti21}.\\
The use of stellar calibrators has also been chosen as the flux calibration method for Ariel. The study of stellar calibrators for Ariel started with the Exoplanet Characterisation Observatory (EChO) mission concept \citep{Tinetti12}, which was Ariel's precursor. The EChO stellar calibrator candidate list was drawn up from 2MASS \citep[Two Micron All Sky Survey,][]{Skrutskie06} and included 537 stars; mainly G dwarfs, but also some F and K dwarfs \citep{echo_calibrators}. These spectral types were chosen for their quiescence \citep{Gautschy96, Ciardi11}. Using the first quarter of data from the Kepler mission \citep[NASA, 2009-2013,][]{Koch10}, \citet{Ciardi11} concluded that G dwarfs are the most stable spectral class, with 80\% of them deemed photometrically stable, followed by K and F dwarfs, at 50\%. The other spectral types have higher variability rates. It was therefore thought that many stars were going to be suitable for the calibration. Kepler however, only observed a small region of the sky, and similar data with a larger sky coverage was not available at that time. So, it was not possible to test stellar variability throughout the sky.\\

Nowadays, this analysis is possible thanks to the TESS mission \citep[Transiting Exoplanet Survey Satellite, NASA, 2018-present,][]{Ricker15}, which has been monitoring bright stars for periods of around 27 days. In July 2020, it completed its two-year-long primary mission, which covered about 75$\%$ of the sky, an area about 400 times bigger than the one covered by Kepler. Its extended mission is still ongoing, collecting additional data and revisiting regions of the sky with longer observing intervals. Time series over timescales of weeks are freely available at Mikulski Archive for Space Telescopes (MAST)\footnote{\url{https://mast.stsci.edu/portal/Mashup/Clients/Mast/Portal.html}}. At the time of writing, more than 7,000 candidate exoplanets have been identified, of which 432 have been confirmed. TESS light curves are ideally suited to identify and characterise stellar calibrators \citep{Ricker15}, e.g. TESS was employed to search for photometric variability in candidate JWST standard stars \citep{Mullally22}.\\

In this paper, we present a flux variability analysis of potential stellar calibrators with TESS data, expanding on the work of \citet{Mullally22}. In Section \ref{ch2}, we describe the method used, from the creation of a preliminary stellar calibrator candidate sample, the TESS light curve data and its employment, the mathematical algorithms and tools used for the flux analysis, to the criteria to create a calibration target list. In Section \ref{ch3}, we show the results of the analysis, including a categorisation and a population analysis of the sample, and the selection of stellar calibrators using the current Ariel flux stability requirement. In Section \ref{ch4}, we discuss the use of different stability requirements for the selection and the limits of TESS data and the presented method. Section \ref{ch5} sums up the most important results of this study.


\section{Methods}\label{ch2}
\subsection{Stellar calibrator candidate sample}\label{sec:cal_starting_sample}
To create a reliable stellar calibrator candidate sample, a pre-selection must be carried out to find stars within the correct magnitude and color ranges relevant to the instrument, in anticipation of stellar instabilities \citep{Gautschy96}. All-sky surveys at optical and near-infrared wavelengths, such as 2MASS and Gaia \citep[e.g. Data Release 3 (DR3),][]{Prusti16, Vallenari22} can be used. Limits are set to identify nearby G, early K, and late F dwarfs, following the flux stability considerations of \citet{Ciardi11}.\\

For this work, the pre-selection was performed with the Gaia DR3 catalogue by setting limits as shown in Table \ref{tab:limits}. The photospheric surface temperature is between 5,000 and 6,000$K$, the surface gravity is between 4.0 and 4.5$\log(cm$ $s^{-1})$. The last limit is biased to nearby stars, also mitigating interstellar reddening effects \citep{Deustua}. In addition, a maximum distance between Earth and the stars of $50pc$ was set, to limit the number of compatibilities when querying Gaia DR3, which only shows the first 2,000 objects. 2,000 objects is a big enough starting sample given our expectations on the number of stable stars and their fairly homogeneous distribution throughout the sky.

\begin{table}[ht]
\centering
\caption{\label{tab:limits} Selection limits for the pre-selection of stellar calibrator candidates applied to the Gaia DR3 catalogue.}   
    \begin{tabular}{l c c }
    \hline
      Parameter & Lower limit & Upper limit\\ 
      \hline\hline
      Surface temperature [K] & 5,000 & 6,000 \\
      Surface gravity [$\log(cm~s^{-1})$] & 4.0 & 4.5 \\ 
      Distance [$pc$] & - & 50 \\
      \hline
    \end{tabular}          
\end{table}

\subsection{TESS data}
For the flux analysis of the stellar calibrator candidates, light curve observations from TESS are used, specifically, the light curve files processed by the Science Processing Operations Center \citep[SPOC,][]{Jenkins16} data processing pipeline developed by the NASA Ames Research Center. The SPOC pipeline provides high-quality background subtraction and correction of known instrumental systematics. TESS data is publicly available on MAST\footnote{This can be queried through the \texttt{lightkurve} \citep{lightkurve} Python package. See \url{https://docs.lightkurve.org}.}.\\

Flux data with a cadence of 2 minutes obtained through Pre-search Data Conditioned Simple Aperture Photometry (PDC SAP) extraction is used. These light curves have undergone a co-trending procedure to remove systematic instrumental effects, as detailed by \citet{smith_pdc_sap}. This procedure follows a Bayesian approach called Maximum A Posteriori by fitting
vectors called Cotrending Basis Vectors and subtracting them from the raw flux data. Therefore, PDC SAP represent the best estimate of the intrinsic stellar variability, with instrumental effects removed. Generally, a further de-trending or filtering of the light curves is performed to smooth the data, but in this work it is assumed that the PDC SAP data quality is good enough for the objective of selecting reliable calibrators, also if this means that some good calibrators might be discarded occasionally because of some remaining instrumental systematics.\\

If data is available in different TESS Sectors, only the first one in chronological order was used. Each TESS Sector has a total length of $\sim$27 days. Following the recommendations of the TESS Archive Manual\footnote{\url{https://outerspace.stsci.edu/display/TESS/2.0+-+Data+Product+Overview}}, the Cadence Quality Flags given by the pipeline are used to remove data points during anomalous events, and/or instrument or spacecraft-related events. In addition, the magnitude in this study is defined as the one adopted in the TESS mission.
\subsection{Lomb-Scargle periodogram}
For analysis and quantification of photometric variability, the Lomb-Scargle periodogram (LSP) is adopted \citep{Lomb76, Scargle82}. The LSP is a well-known algorithm to detect periodic variability in irregularly sampled data in a limited time domain \citep{vanderPlas18}. Although TESS observations are aimed to be uniformly spaced with a 2-minute cadence, data shows a spread in cadence in the order of tens of milliseconds, originating in the CCD readouts. In addition, data gaps up to minutes are common due to anomalies such as the presence of the Earth in the field of view, re-orientation of the spacecraft, cosmic ray impacts, processing errors, or other phenomena.\\
The LSP enables the study of the frequency content of the signal and the identification of potential periodic variability. It fits groups of simple sinusoids at different candidate frequencies and estimates the goodness of fit by minimising the residuals. The LSP approach has been adopted in different studies on stellar rotation or pulsations, including variability studies of JWST spectrophotometric standards \citep{Ma22, Mullally22}.\\
The frequency range of the LSP should be set from the inverse of the total signal duration, so that the signal can complete at least one full sinusoidal oscillation during its time duration, to the Nyquist frequency. Since TESS data is nearly uniformly spaced, we assume the regular Nyquist frequency is a good approximation. In addition, to not miss power peaks, the frequency range should be sampled finely enough. So, the spacing must be at least equal to the expected widths of the LSP power peaks, or smaller \citep{vanderPlas18}. To sum up:
\begin{equation}
    \begin{split}
        f_{min}&=\frac{1}{T}\\
        f_{max}&=\frac{f_{samp}}{2}\\
        \Delta f&=\frac{1}{nT}, n\geq1
    \end{split}
\end{equation}
where $f_{min}$ and $f_{max}$ are the minimum and maximum frequency, respectively, and $\Delta f$ is the frequency spacing. $T$ is the signal duration in time units, $f_{samp}$ is the sampling frequency or cadence of the data, and $n$ is a multiplicative factor that must be greater or equal to $1$.\\

Finally, to have a straightforward relation between the peaks in the LSP power spectrum and the oscillation amplitudes in the signal, the power is scaled into power amplitude with a proportion derived from \citet{Kjeldsen92} and \citet{Kjeldsen94}:
\begin{equation}
    A(f)=\sqrt{\frac{4}{N}P_S(f)}
\end{equation}
where $P_S(f)$ is the power spectrum as a function of frequency $f$, $A(f)$ is the power amplitude spectrum as a function of frequency, and $N$ is the total number of samples in the time series.

\subsection{Statistical quantities of interest}
A useful quantity to characterise the photometric variability of a star from its light curve is the flux dispersion, $\sigma_{Mdn}$, around the median flux value, ${Mdn}$. This is defined as:
\begin{equation}
    \sigma_{Mdn}=\sqrt{\frac{\sum{(x_n-Mdn)}}{N}}
\end{equation}
where $x_n$ represents the time series points, $Mdn$ is the median, and $N$ the total number of points. 
Flux dispersion alone is not enough to assess stellar flux variability, as it depends on some stellar properties like the magnitude. So, the reduced chi-squared, $\chi_v^2$, is inspected; a unitless quantity that describes how much the data varies on top of the uncertainty in the measurements. The uncertainty can be considered to be the instantaneous noise in the data points. This is the flux error provided in the TESS data files. The reduced chi-squared is defined as:
\begin{equation}
    \chi_v^2=\frac{\chi^2}{K_f}=\frac{1}{N-1}\sum_{n=1}^N\left(\frac{x_n-Mdn}{\sigma_n}\right)^2
\end{equation}
where $\chi^2$ is the chi-squared, and $\sigma_n$ is the instantaneous uncertainty in the data points. $K_f$ corresponds to the degrees of freedom i.e. the difference between the number of data points and the fitted parameters. In this case, the number of data points is $N$ and the number of fitted parameters is 1 since the reference around which the reduced chi-squared is defined is constant, i.e. the median of the time series.\\

The larger $\chi_v^2$, the larger the flux variation with respect to the instantaneous noise. Following \citet{Ciardi11}, it can be assumed that a star is very variable if $\chi_v^2\geq100$ (excess flux dispersion of $\sim10$ times the flux uncertainty), significantly variable if $10\leq\chi_v^2<100$ (excess flux dispersion of $\sim3$ times the flux uncertainty), just barely variable if $2\leq\chi_v^2<10$ (excess flux dispersion of $\sim1.5$ times the flux uncertainty), and not variable if $\chi_v^2<2$, with respect to the flux error. In the latter case, the flux variation is not significant compared to the measurements' error. 

\subsection{Selection criteria}\label{sec:sel_criteria}
After the pre-selection, a procedure composed of three steps is applied to select photometrically stable stellar calibrators from a starting candidate sample. The selection criteria are as follows:
\begin{enumerate}
    \item The LSP power amplitude as a function of frequency is lower than the desired flux stability level as a function of frequency. Mathematically, $A(f)<FSL(f)/2$, where $FSL(f)$ is the desired peak-to-peak flux stability level as a function of frequency within the LSP.
    \item The reduced chi-squared is lower than 2. Mathematically, $\chi_v^2<2$.
    \item The stellar calibrator candidate does not host any confirmed or candidate exoplanets.
\end{enumerate}

\begin{figure*}[ht]
\centering
\includegraphics[width=15.5cm]{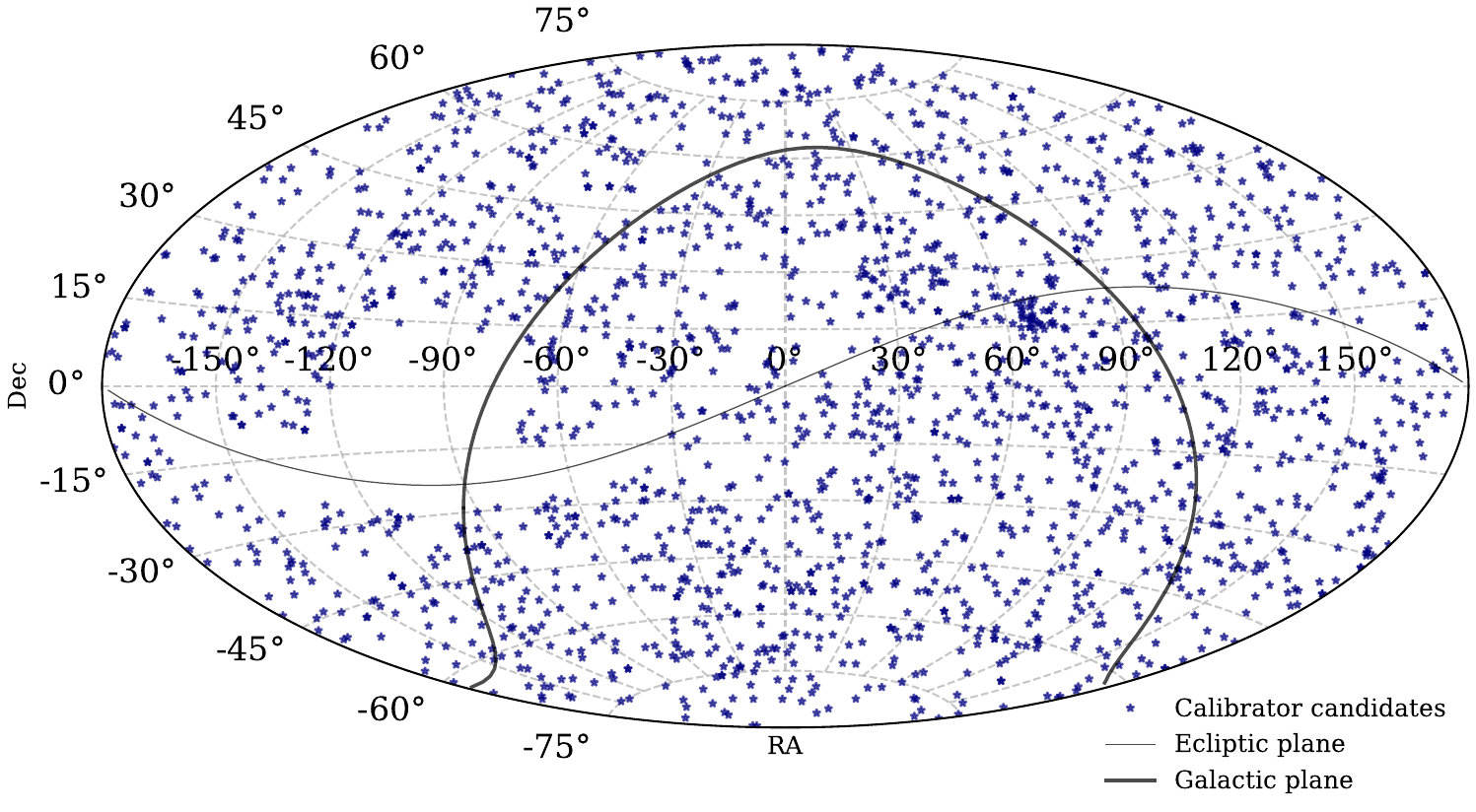}
    \caption{Positions in the sky of the stellar calibrator candidates from the Ariel starting sample in the International Celestial Reference System (ICRS) and at epoch J2000.}
    \label{fig:candidates_sky}
\end{figure*}

The first step uses the LSP power amplitude spectrum to remove the stellar calibrator candidates showing a periodic photometric variability larger than the desired flux stability level from the starting sample. It is not desired to have oscillation components of amplitude larger than the desired flux stability level, which is usually defined for a certain time scale, i.e. it is dependent on the frequency.\\

From \citet{Ciardi11} a star is stable when the excess flux dispersion is lower than 1.5 times the flux uncertainty, i.e. the flux variation is not significant compared to the measurement error. So, the reduced chi-squared step removes the stars with a high flux dispersion.\\

The third step removes the stars that are known to host at least one exoplanet. To be on the safe side also stars with transiting planetary candidates identified by TESS, called TESS Project Candidates, should be removed from the sample. This is because the flux of stars hosting exoplanets might be affected by dips or other flux variations due to exoplanet transits or other phase curve effects. This criterion is the least stringent, as planets that are small or far away from their star might not significantly affect the observed flux. In case of a shortage of stellar calibrators not hosting exoplanets, very stable host stars could still be taken into consideration as calibrators when observed outside of transit times. However, to do so, a careful evaluation of each host star would be necessary. For simplicity, in this study, all host stars are removed from the sample both for transiting and not transiting exoplanets.\\

The stellar calibrator candidates that do not pass these criteria should not be further considered, as they appear to vary too much photometrically. The other stars instead would be, in first approximation, mainly dominated by noise, and seen as stable, i.e. good stellar calibrator candidates.

\section{Results}\label{ch3}
\subsection{Pre-selection}
To create a starting candidate sample, a list was constructed from the method in
\ref{sec:cal_starting_sample}. This resulted in 1,907 candidates. Subsequently, it was combined with the original EChO calibrator candidate list. After removing duplicates, this pre-selection resulted in a sample containing 2,244 stars. For 307 stars, no TESS data or TESS data of sufficient quality was available, so they were removed from the sample. Most of the removed stars are near the galactic centre and were not included in the TESS survey due to crowding. Figure \ref{fig:candidates_sky} shows the positions of the remaining 1,937 calibrator candidates in the sky.\\

\begin{figure*}[th]
\centering
\includegraphics[width=17.5cm]{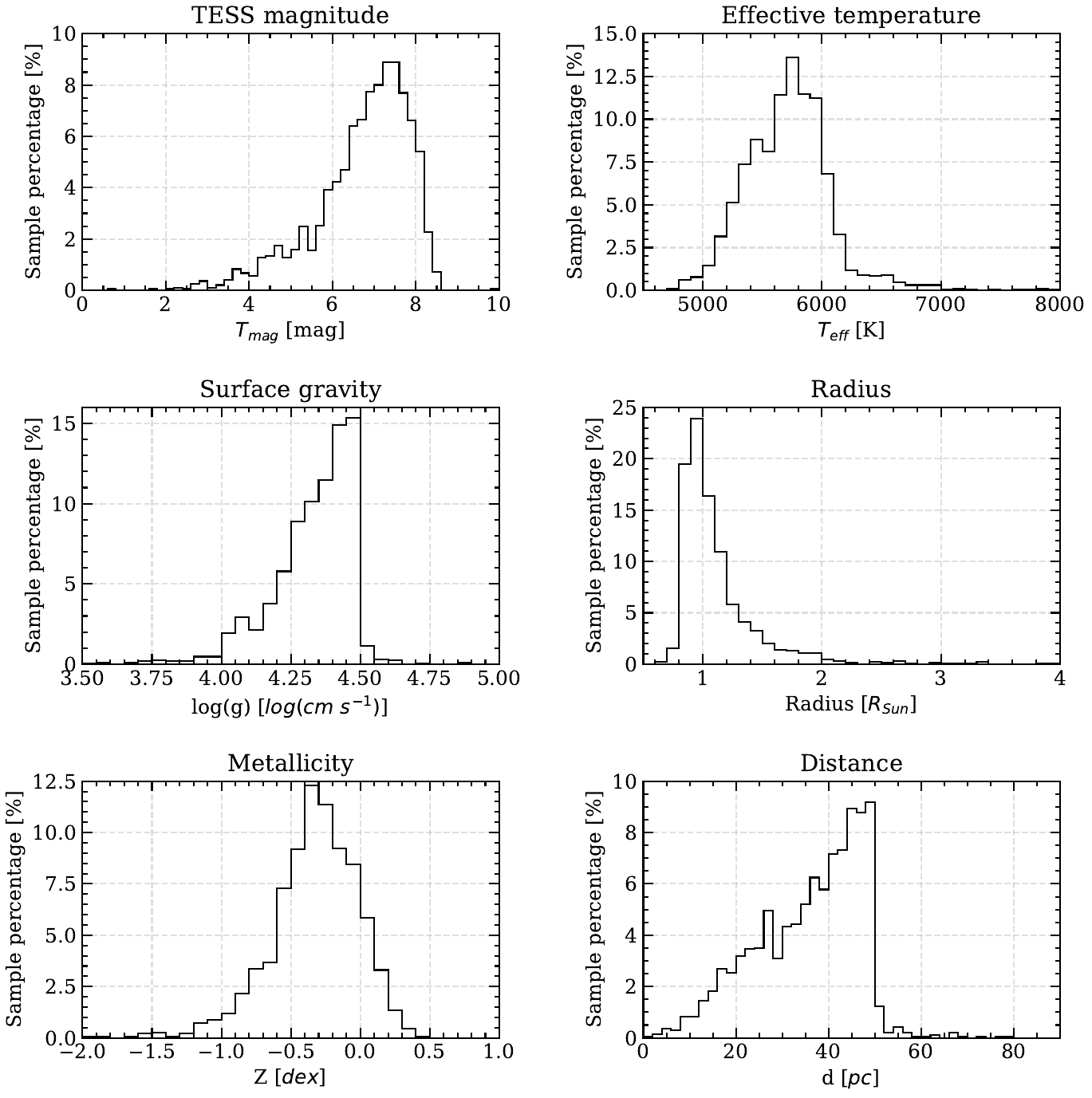}
    \caption{Distributions of the starting Ariel stellar calibrator candidate sample properties including magnitude, effective temperature, surface gravity, radius, metallicity, and distance between star and observer.}
    \label{fig:candidates_hist}
\end{figure*}

The histograms in Figure \ref{fig:candidates_hist} show the distributions of selected stellar properties of the candidate sample, including the magnitude, effective temperature, surface gravity, radius, metallicity, and distance. The magnitude, together with surface temperature, radius, and surface gravity, gives us an indication of the stellar spectral class. Stellar age, tracked by metallicity, is uncertain and not always available but it is interesting to inspect flux variability, as younger stars tend to be more active and variable. 

\subsection{Categorisation}
From the LSP analysis, five main stellar categories can be identified within the calibrator candidate sample. These categories are arbitrarily defined to give an idea of the composition of the sample to the reader. Figures \ref{fig:cat_ex_1} and \ref{fig:cat_ex_2} show examples of the light curves and associated LSP of the different types of behaviours that were identified. The categories are defined as follows:
\begin{enumerate}
    \item Stars with transit events, including stars hosting transiting exoplanets and stars in eclipsing binary systems. They are found through cross-match with the NASA Exoplanet Archive\footnote{\url{https://exoplanetarchive.ipac.caltech.edu/index.html}} and the TESS eclipsing binary stars catalogue \citep{Pra22}, respectively. This catalogue does not cover the most recent TESS sectors. The list of known exoplanets from the NASA Exoplanet Archive was downloaded on the 22nd of May 2023 and so it only includes exoplanets or candidate exoplanets known at that time.
    \item High-frequency pulsating stars. They possess an LSP power peak greater than or equal to $50ppm$ at a frequency $f>1d^{-1}$.
    \item High-amplitude rotating stars. They possess an LSP power peak greater than or equal to $100ppm$ at a frequency $f\leq1d^{-1}$.
    \item Low-amplitude rotating stars. They possess an LSP power peak greater than or equal to $50ppm$, but lower than $100ppm$, at a frequency $f\leq1d^{-1}$.
    \item Low variability stars. These are valid stellar calibrator candidates. The LSP power is lower than $50ppm$ at all frequencies within the LSP.
\end{enumerate}

The limits imposed on the amplitude are completely arbitrary, and a rigorous identification of stable stars will be performed in the next section. The frequency limit to divide stars between pulsating and rotating types instead is based on typical timescales for such stars. For instance, pulsating variables like $\delta$ Scuti or $\gamma$ Doradus pulse at timescales of 18min-8hrs and \~1 day respectively (\citet{delta_scuti}, \citet{gamma_doradus}), while stars whose light curves are mainly modulated by rotation usually rotate at timescales between 1 and 25 days \citep{stellar_book}. Hence, a frequency of 1$d^{-1}$ is used to divide the two types.

\begin{figure*}[th]
\centering
    \includegraphics[width=17.5cm]{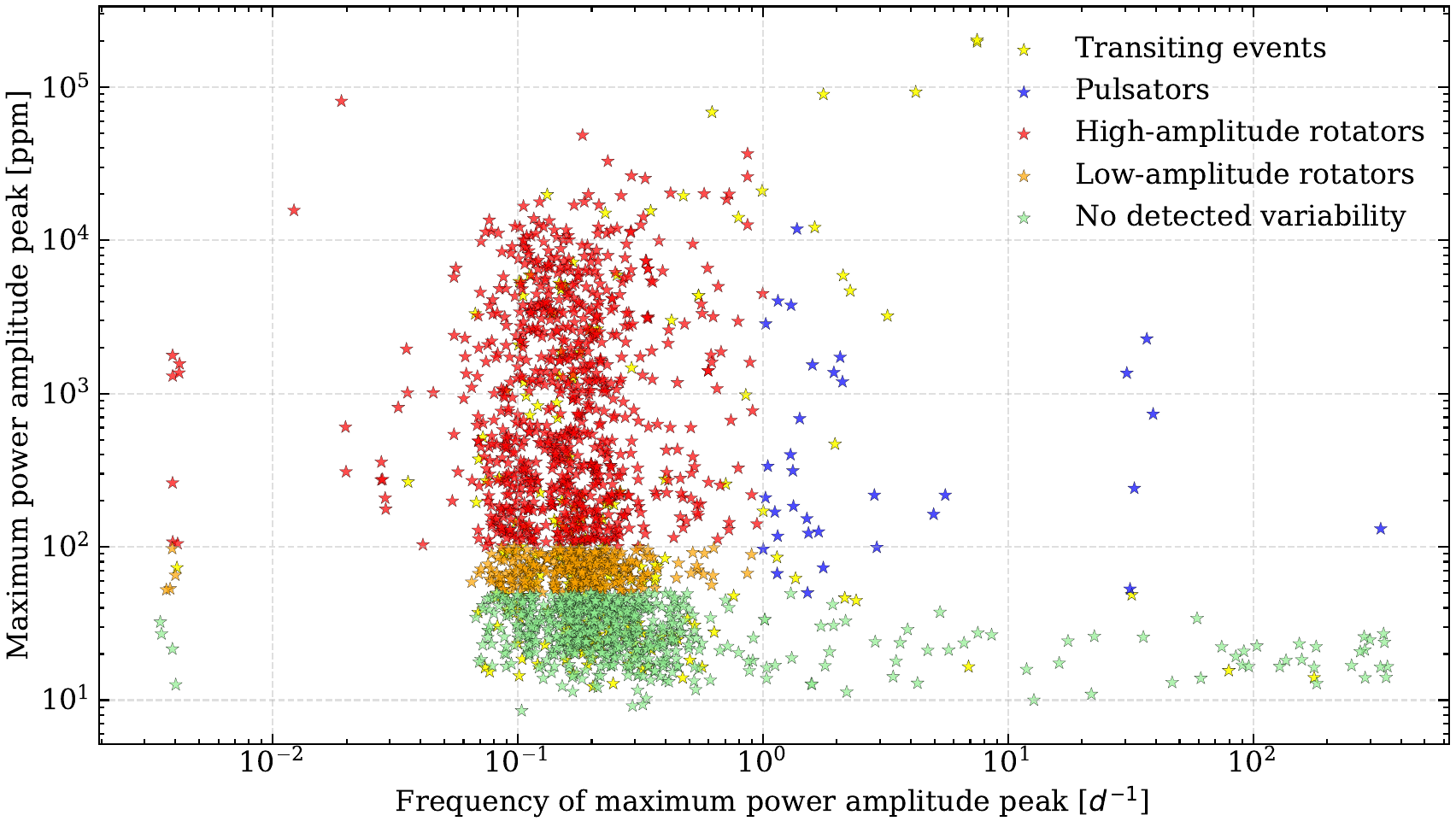}
    \caption{Distribution of the stars within the starting Ariel stellar calibrator candidate sample, divided by stellar category, in a frequency-power plot. The y-axis represents the maximum LSP power peak and the x-axis represents the frequency at which it occurs.}
    \label{fig:cat_type}
\end{figure*}

\begin{table*}[th]
\centering
\caption{\label{tab:categories} Number of stars from the calibrator candidate sample belonging to each defined stellar category. Stars with mixed features are arbitrarily assigned to one category only. The distinction between the low-amplitude rotating stars and stars without variability is given as a first approximation by arbitrarily setting a maximum allowed LSP power amplitude peak of $50ppm$ for stars to be considered non-variable.}   
    \begin{tabular}{l c c c c}
    \hline
      Category & f($P_{max}$) [$d^{-1}$] & $P_{max}$ [$ppm$] & Number of stars & Sample percentage\\ 
      \hline\hline
        Transit events & any & any & 212 & 10.95\% \\
        Pulsating stars & $>1$ & $\geq50$ & 33 & 1.70\% \\
        High-amplitude rotating stars & $\leq1$ & $\geq100$ & 801 & 41.35\% \\
        Low-amplitude rotating stars & $\leq1$ & $\geq50$ and $<100$ & 282 & 14.56\% \\
        Low variability & any & $< 50$ & 609 & 31.44\% \\
        \hline
    \end{tabular}          
\end{table*}

Stars can exhibit multiple features among these categories, e.g. they can possess rotation features and transiting events, or rotate and pulsate at the same time. Table \ref{tab:categories} shows how many stars from the calibrator candidate sample belong to each category. Stars with different features are assigned to a single category as follows:
\begin{itemize}
    \item Stars possessing transit events combined with any other flux variability feature are only included in category A (stars with transit events);
    \item Stars with both rotation and pulsation features belong to one or the other group depending on the frequency at which the highest LSP power peak occurs. If the highest peak occurs at a frequency $f>1d^{-1}$, the star is included in category B (pulsating stars), and otherwise in category C or D (rotating stars).
\end{itemize}

Following this categorisation, high-amplitude rotating stars make up the majority of the sample ($\sim$41\%), followed by stars with low variability ($\sim$31\%), and low-amplitude rotating ones ($\sim$15\%). Stars affected by transiting events are $\sim$11\% of the sample, and pulsating stars make up the smallest category, being less than 2\% of the sample.

Figure \ref{fig:cat_type} shows the distribution of the stars in the calibrator candidate sample as a function of their maximum LSP power amplitude peak and the frequency at which it occurs. The high and low-amplitude rotating stars and stars with low variability are clustered within a frequency range between $f=0.07d^{-1}$ and $f=0.5d^{-1}$, corresponding to stellar rotation periods between 2 and 14 days. The cluster appears to be bimodal, centered at $f=0.1d^{-1}$ and $f=0.2d^{-1}$. These correspond to rotation periods of 5 and 10 days, respectively. The stars with no detected variability are spread in the lower right part of the plot, where frequencies are higher. This is because, as noise becomes the dominant factor in the light curve, the highest peak is a sum of random noise components that can happen at any frequency. Some stars have maximum power peaks at very low frequencies ($f\sim0.006d^{-1}$). Such stars show a drift in flux, which can either have a physical origin (i.e. a flux variation on a really long time scale) or an instrumental one (i.e. systematics that were not properly corrected). By definition, pulsating stars are found in the right part of the plot at high frequencies. Stars undergoing transit events populate the full plot without being correlated to specific power or frequency values.

\begin{figure*}[th]
\centering
    \includegraphics[width=17.5cm]{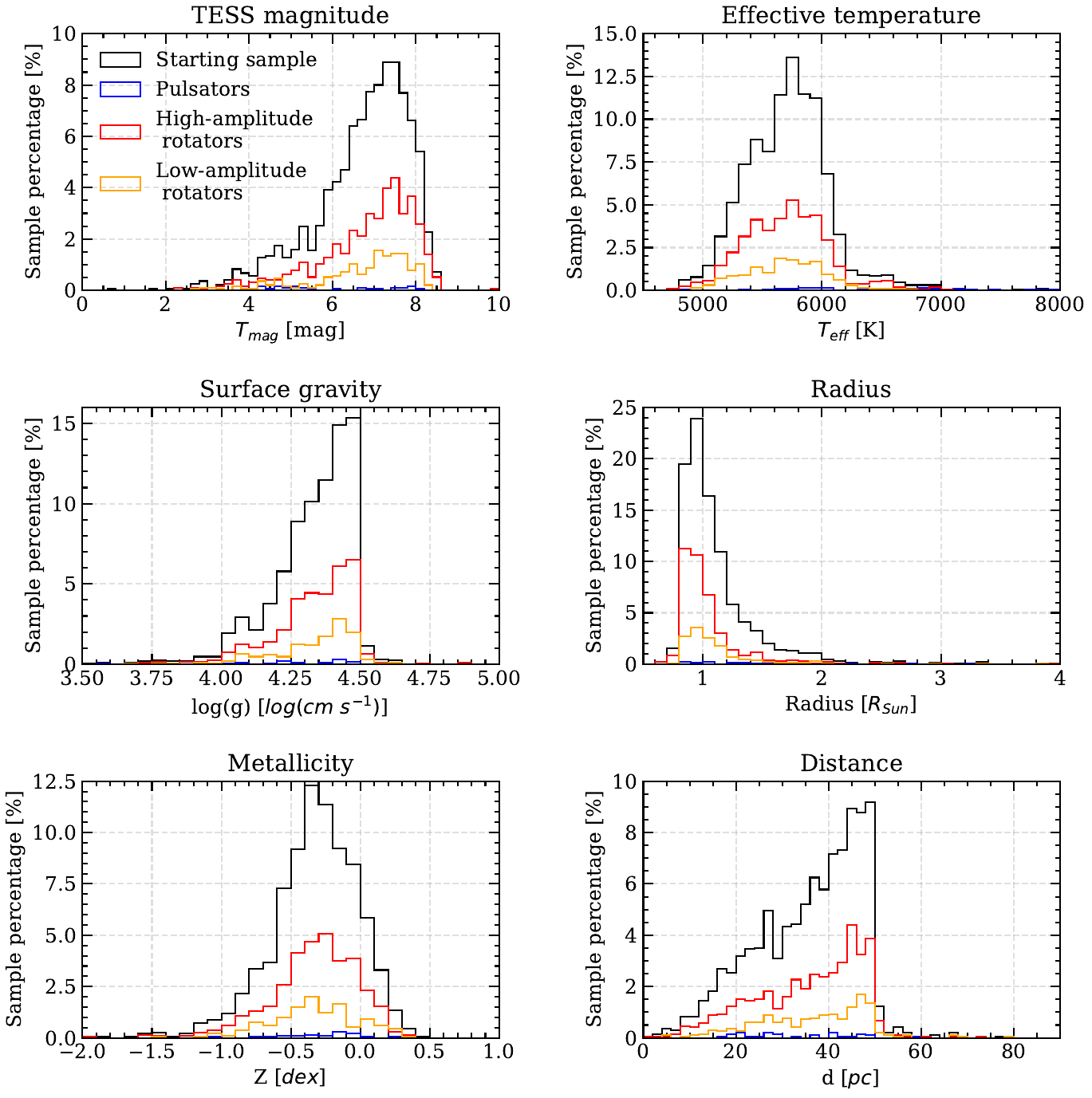}
    \caption{Distributions of the starting stellar calibrator candidate sample properties after the categorisation.}
    \label{fig:cat_pop}
\end{figure*}

Figure \ref{fig:cat_pop} shows the distributions of stellar properties for three categories. The high and low-amplitude rotating stars follow almost exactly the distribution of the whole sample for each property. This is expected since these stars make up the majority of the sample. The pulsating stars instead, only cover a specific temperature range: They are mainly found at the highest temperatures of the sample (7,000-8,000$K$), with some others around 5,500-6,000$K$.

\subsection{Calibrators selection}
The selection criteria described in Section \ref{sec:sel_criteria} are applied with the Ariel flux stability requirement value of 100$ppm$ over 6 hours. Since this requirement is peak-to-peak, it is not desired to have oscillation components of amplitude larger than 50$ppm$ at a time scale of 6 hours or lower, i.e. for $f\geq4d^{-1}$, in the LSP. Stellar flux time series do not have perfect sinusoidal shapes at specific frequencies but rather possess multiple components that can add up, making the flux possibly fluctuate more than 100$ppm$. The maximum allowed amplitude peak value of 50$ppm$ is extended to the whole frequency range within the LSP, which does not solve completely the problem of beating frequencies, but it eases it partially. The third criterion is performed through a cross-match of the starting calibrator candidate sample with the all the known or candidate exoplanet hosts from the NASA Exoplanet Archive’s Planetary Systems Table \citep{NExScI20}, downloaded in date 22nd of May 2023. \\

Table \ref{tab:selection} shows how many stars from the calibrator candidate sample pass the selection criteria given in Section \ref{sec:sel_criteria}. Criteria are applied in consecutive order, i.e. the stars passing a criterion are given as input to the following criterion.\\

\begin{table}[th]
\centering
\caption{Number of stars from the calibrator candidate sample passing the selection criteria. The criteria are applied in the order shown and in a consecutive way.}
    \begin{tabular}{l c c }
    \hline
        Criterion & Stars passing & Percentage passing \\
        \hline \hline
        1. LSP & 703 & 36.29\% \\
        2. $\chi_v^2$ & 488 & 25.19\% \\
        3. Host star & 430 & 22.20\% \\
        \hline
    \end{tabular}
    \label{tab:selection}
\end{table}

\begin{figure*}[th]
\centering
    \includegraphics[width=15.5cm]{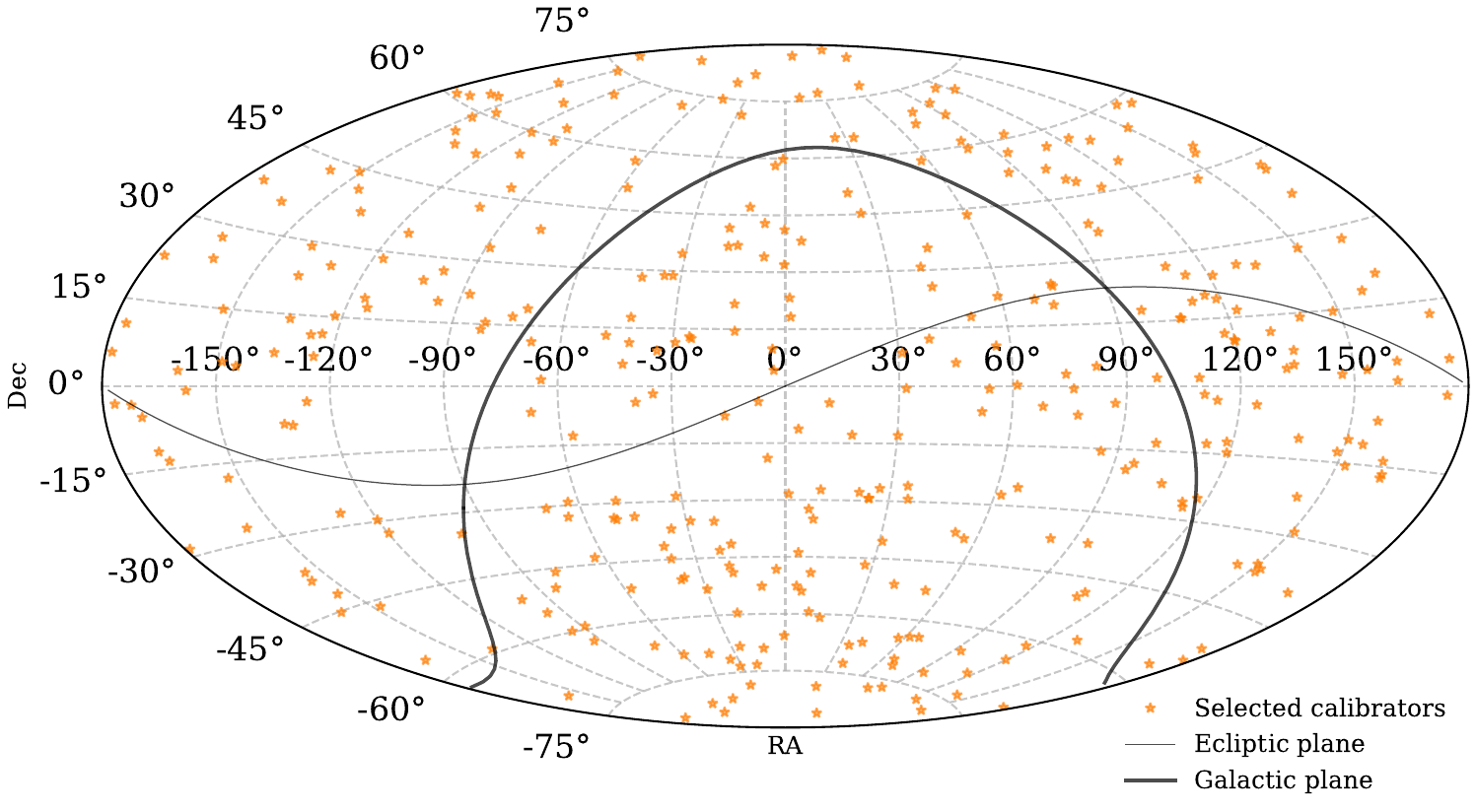}
    \caption{Positions in the sky of the selected stellar calibrators in the ICRS and at epoch J2000.}
    \label{fig:sky_selected}
\end{figure*}

The first step removes stars possessing macro-variability, i.e. periodic flux variations with amplitudes that are too large for the desired stability requirement. These include most pulsating stars, all high-amplitude rotating stars, and all eclipsing binaries. Almost half of the sample is removed in the first step of the selection. The second step filters the stars with a large excess flux dispersion compared to the measurements' uncertainty. These include low-amplitude rotating stars and stars that possess multiple variation features. Finally, the third step removes known or candidate stars hosting exoplanets.\\

In total, 430 stars pass the full selection procedure: $\sim$22.2\% of the starting sample. This is the selected Ariel stellar calibrator list. The stars in this list do not possess a large periodic variability in their fluxes, an excess dispersion of more than 1.5 times the measurement uncertainties, and do not host known or candidate exoplanets. This makes them good potential stellar calibrators.\\

\begin{figure*}[th]
   \centering
   \includegraphics[width=17.5cm]{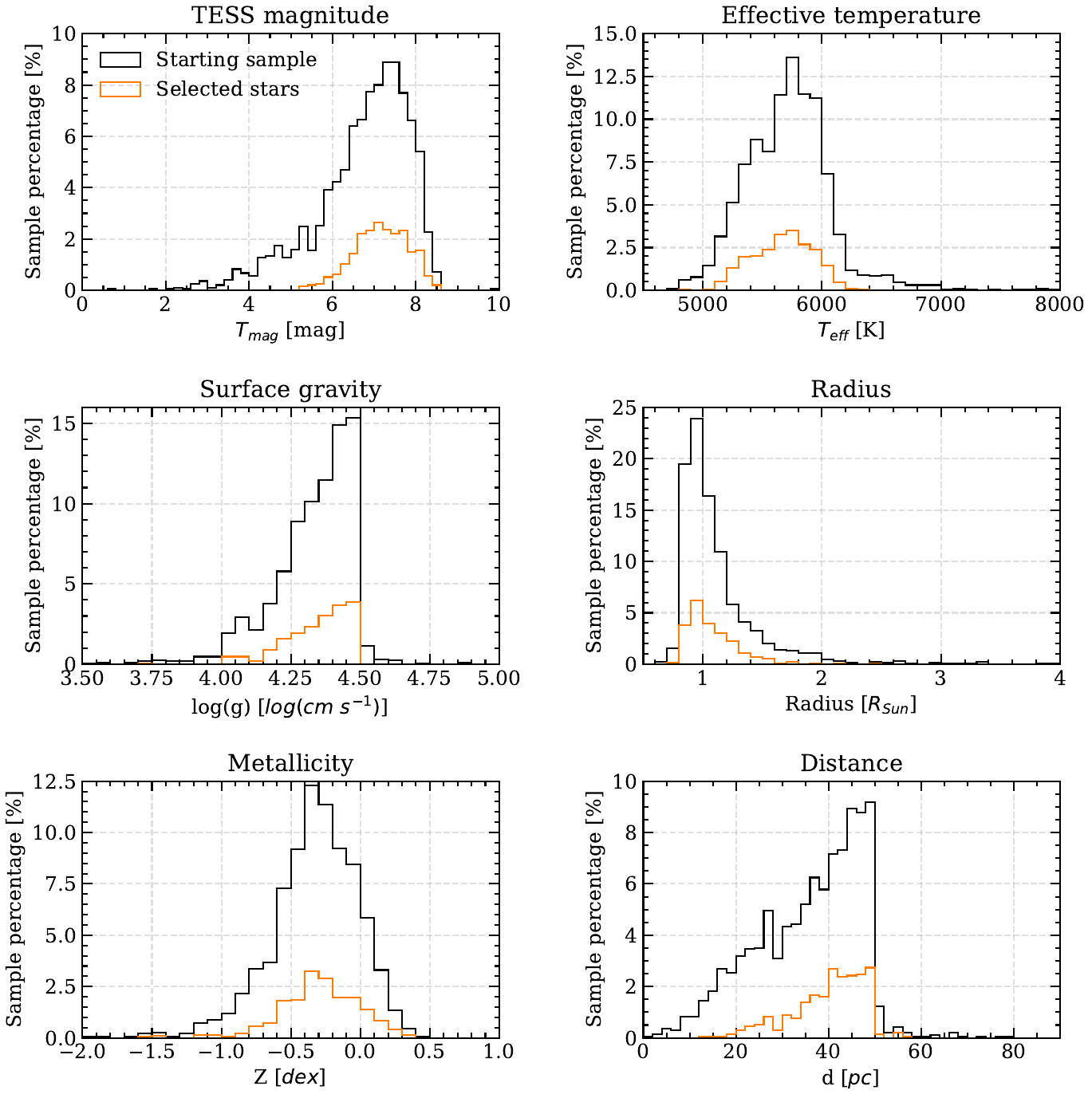}
    \caption{Distributions of the properties of the selected stellar calibrators.}
    \label{fig:pop_sel}
\end{figure*}

Figure \ref{fig:sky_selected} presents the positions of the selected stellar calibrators in the sky. These appear to be distributed homogeneously. A homogeneous distribution is desired to minimise slewing of the spacecraft \citep{Petralia17}. Figure \ref{fig:pop_sel} shows the distributions of the stellar properties of the selected stellar calibrators with respect to the starting candidate sample. The most important conclusion is seen in the stellar magnitude: No stars brighter than 5$mag$ pass the selection. This is likely due to the inherent noise level of TESS itself. Bright stars possess lower measurement uncertainties, and flux variability is more easily identifiable. Dimmer stars are more likely to pass the selection procedure, effectively hiding flux variability within the noise.\\

The stars passing the selection only possess effective temperatures in the range of 5,000-6,300$K$. For surface gravity, radius, metallicity, and distance between observer and star values span almost the same ranges of the starting sample.\\

The distributions of the selected stellar calibrators closely resemble the starting candidate sample in shape for all parameters, indicating that, once the correct range of values for each stellar property is identified, the probability of finding a stable star inside such ranges remains approximately constant at $\sim$22\%. The only exception is the magnitude, with dimmer stars passing the selection criteria more often than brighter ones. As previously mentioned, this bias is hypothesised to originate from the noise level in the TESS light curves. In addition, this result was not anticipated for metallicity, since stars with low metallicity (assumed to be older, i.e. less active) were expected to have a higher probability of being stable.

\subsection{Variability study}
With the statistical quantities calculated for each stellar light curve, a variability study of the calibrator sample is performed. The results are compared with the study by \citet{Ciardi11} to determine if the considerations on stellar flux stability drawn from the restricted region observed by Kepler can be extended to the whole sky.\\

\begin{figure*}[th]
   \centering
   \includegraphics[width=15.5cm]{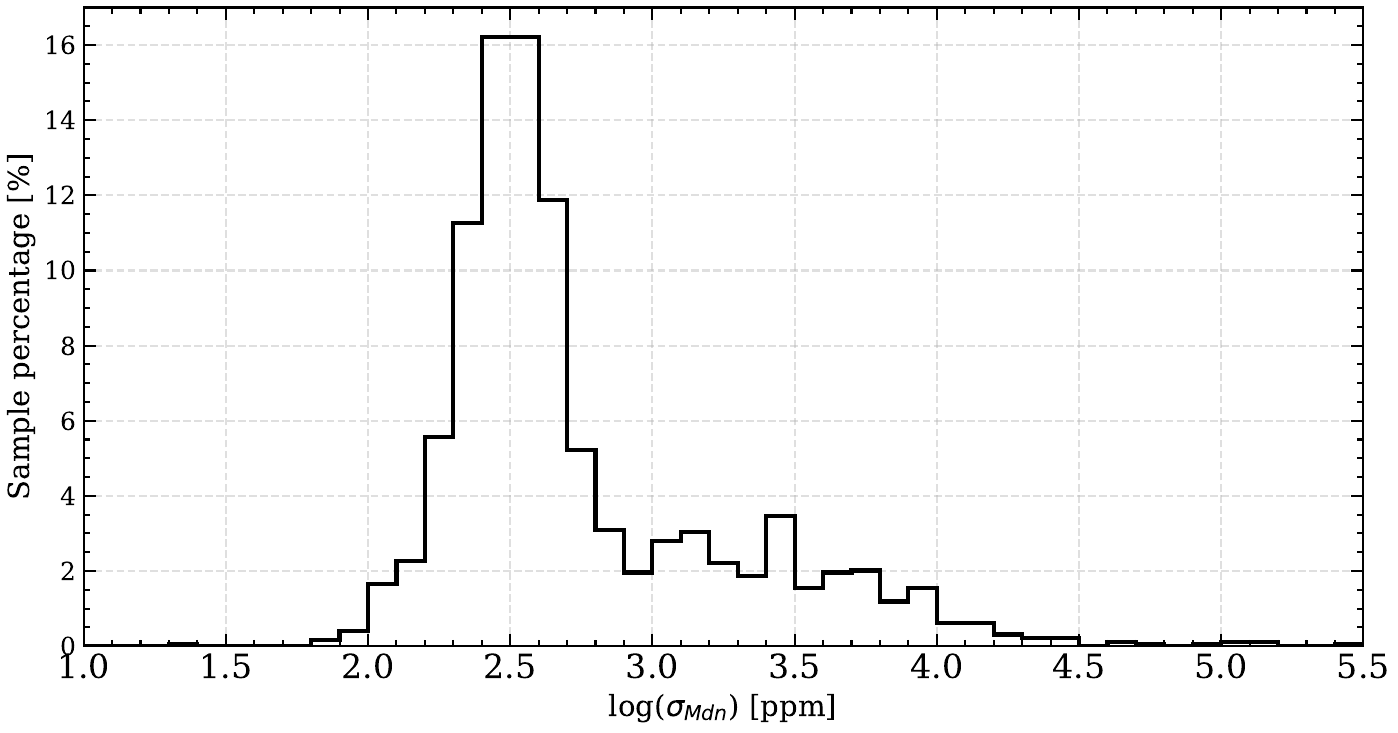}
    \caption{Distribution of the logarithmic flux dispersion around the median of the stars in the calibrator candidate sample.}
    \label{fig:ciardi7}
\end{figure*}

\begin{figure*}[th]
   \centering
   \includegraphics[width=15.5cm]{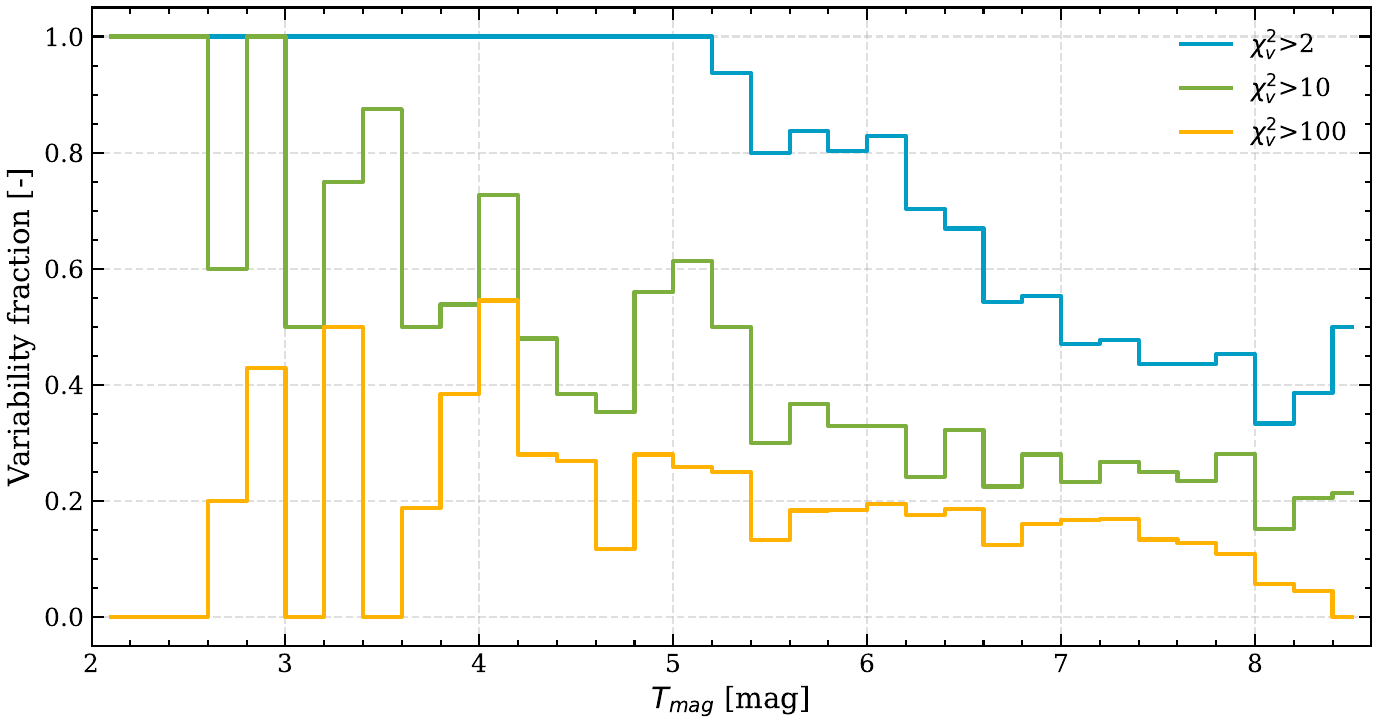}
    \caption{Variability fractions of the stars in the calibrator candidate sample as a function of magnitude. Following \citet{Ciardi11}, the blue curve represents the fractions of stars with $\chi^2_v>2$, the green curve the fractions of stars with $\chi^2_v>10$, and the orange curve the fractions of stars with $\chi^2_v>100$.}
    \label{fig:ciardi10}
\end{figure*}

Figure \ref{fig:ciardi7} is equivalent to the panel of Figure 7 of \citet{Ciardi11} discussing G dwarfs. It displays the distribution of the computed flux dispersion around the median of the stars in the starting candidate sample. A bimodal distribution is visible, with a main group of stars in the range $100<\sigma_{Mdn}<1,000ppm$ and a second, smaller group in the range $1,000<\sigma_{Mdn}<10,000ppm$. $\sim$80\% of stars in the starting sample belong to the lower dispersion group, and around $\sim$20\% to the higher dispersion one. In the study by \cite{Ciardi11}, the two groups include $\sim$90\% and $\sim$10\% of their starting sample, i.e. the G dwarfs from the first quarter of data from Kepler. The percentages are similar between the two studies, and their difference could be due to a bias related to the limited size of the sample used in this study (around 2,000 stars) compared to the sample of \cite{Ciardi11} (around 66,000 stars).\\

Figure \ref{fig:ciardi10} shows the variability fractions of the stars in the starting sample for different variability levels defined by \citet{Ciardi11} and is equivalent to the panel of Figure 10 by \citet{Ciardi11} discussing G dwarfs. These fractions show a strong dependency on stellar magnitude, i.e. no stars brighter than $5mag$ are stable and, as the stars get dimmer, the variability fractions decrease accordingly. This decreasing trend is smooth in \citet{Ciardi11} and caused by the degraded photometric precision of the instruments for dimmer stars. Figure \ref{fig:ciardi10} corroborates this, with only minor fluctuations attributed to the limited size of this sample. Moreover, this trend was already identified in the magnitude panel of Figure \ref{fig:pop_sel}.\\

It is possible to conclude that the general distributions and trends found by \citet{Ciardi11} for the small area of the sky covered by Kepler are confirmed in this study with data with the wider TESS sky coverage.

\section{Discussion}\label{ch4}
\subsection{Change in flux stability requirement}

\begin{figure*}[th]
    \centering
    \includegraphics[width=15.5cm]{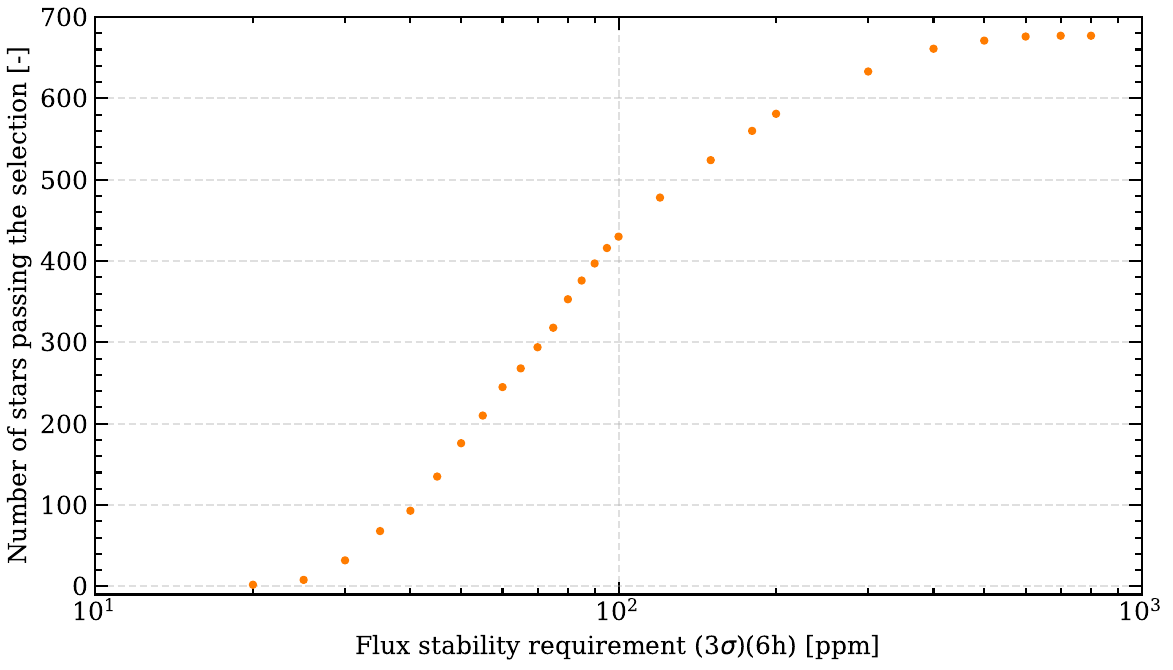}
    \caption{Percentage of stars from the starting sample passing the selection procedure for different values of the flux stability requirement.}
    \label{fig:sens_analysis}
\end{figure*}

The method presented can be adapted to analyse stellar calibrator candidates for other space missions or space mission concepts. The current Ariel flux stability requirement itself could be subject to change in the future. So, it is useful to estimate the percentage of potential stellar calibrators as a function of the required stability level. It is possible to estimate this by changing the maximum LSP power peak value in the first criterion of the selection procedure to the desired flux stability value. The second and third selection criteria are not changed.\\

There are two limitations to the flux analysis of stellar calibrator candidates. First, all stars are inherently rotators, most falling in the high and low-amplitude categories defined above. One must thus assume that all stars are inherently variable, making a flux stability requirement of zero physically impossible. Second, all measurements are limited to either the photon noise limit or the instrumental noise limit, whichever is larger. In this study, TESS is used, which is discussed later.\\

Figure \ref{fig:sens_analysis} presents the percentage of stars from the starting stellar calibrator candidate sample passing the selection procedure for different values of the flux stability requirement. As expected, the percentage increases as the requirement value also increases. At $\sim$300$ppm$, a plateau is reached, counting $\sim$35\% of the stars in the starting candidate sample passing the selection. When the stability requirement is relaxed to 300$ppm$ or higher, the dominant criterion in the selection procedure is the $\chi_v^2$ criterion. No matter how much the flux stability requirement is relaxed, no new sources pass the selection because their flux dispersion is big compared to the flux uncertainty, i.e. they fail to pass the $\chi_v^2$ criterion of the selection procedure. \citet{Ciardi11} found that $\sim$80\% of the analysed G dwarfs possessed a $\chi_v^2$ computed on the full light curve duration lower than 2, i.e. were considered not variable. In this study, only $\sim$35\% of the stars possessed a $\chi_v^2$ less than 2. The reason for this disagreement with the study by \citet{Ciardi11} is currently not known, but is possibly related to the difference in Kepler and TESS sensitivity and stability, as well as the data reduction steps taken in either pipeline. A detailed analysis, although interesting, is complicated given the lack of common stars between this study and the work of \citet{Ciardi11}, and is considered to be beyond the scope of this paper.




\subsection{TESS and LSP limitations}
As previously shown in Figure \ref{fig:pop_sel} and \ref{fig:ciardi10}, the brightest stars do not pass the selection procedure because their variation with respect to the instantaneous noise in the data is large. Specifically, the stars are always seen as too variable when brighter than 5$mag$. TESS's photometric uncertainties are dominated by pointing jitter, with a systematic error floor at 60$ppm$ over 1 hour. This would translate to a noise floor of $\sim$25$ppm$ over 6 hours. For this reason, it is not possible to infer anything about the variation of any star below this level. In fact, Figure \ref{fig:sens_analysis} shows that no stars are selected if the flux stability requirement is too low.\\

On top of the pointing jitter noise, the stellar noise component increases as stars get dimmer, making the total noise increase with magnitude. This means that dimmer stars passing the selection procedure are likely to also vary, but their variation is hidden within the noise. So, underlying structured signals could still be present in the selected stellar calibrators, as well as systematics that were not completely filtered by the pipeline. Moreover, only one TESS Sector for each star was used for the analysis carried out in this study. Different stars were observed in different sectors, which possess a slightly different quality of the data.\\

Stellar calibrators selected with the method presented in this study can thus be considered the most promising stellar calibrators for Ariel. Although this procedure works really well to identify variable stars that are bad calibrator candidates, it is not sufficient to ultimately identify the final list of calibrators and prove their stability within a desired stability level. While the stars not passing this selection are not suitable calibrators, further study on the flux variability of the stars that pass the selection is needed. This can be done by analysing other TESS Sector observations when available, or with targeted future observations, possibly employing observatories with higher precision, e.g. PLATO \citep[PLAnetary Transits and Oscillations of stars, ESA, 2026-,][]{Rauer24}. These stars should be monitored in time and closer to the launch date of the specific mission of interest to make sure their behaviour remains stable in time, as well as to monitor stellar variability features that are short-lived, e.g. coronal mass ejections (CMEs) and flares.

\section{Conclusions}\label{ch5}
Flux stability is a pivotal element to ensure the instrumental accuracy required to perform transit spectroscopy of exoplanetary atmospheres. Among the different flux calibration methods, the use of stellar calibrators acting as external reference sources is particularly promising.\\

In this study, a method was presented to pre-select and carry out a flux variability analysis of stellar calibrator candidates using light curve data from TESS. By computing the Lomb-Scargle periodogram and the reduced chi-squared associated with the stellar light curves, it is possible to analyse the flux variability of the stellar candidate sample. A selection procedure to identify the best stellar calibrators was presented, using the Ariel flux stability requirement of 100$ppm$ $(3\sigma)$ over 6 hours, and a variability population analysis of the sample was carried out.\\

The starting sample mainly includes G dwarfs, distributed all over the sky with the exception of the galactic centre. Following the selection, the most important conclusions of this study include:
\begin{itemize}
    \item The selected stellar calibrator population does not show correlations with stellar properties. Once the correct value ranges for stellar parameters like effective temperature, surface gravity, and similar, are identified, there is a $\sim$22\% probability of finding stable stars within those ranges, with the exception of stellar magnitude. Dimmer stars pass the selection more easily than brighter ones; a bias that is probably originated from the TESS noise level, which increases as stars get dimmer. This causes the flux variation of dimmer stars to be hidden within the inherent TESS noise.
    \item General trends from the stellar variability study by \citet{Ciardi11}, which used the first quarter of data from Kepler, are confirmed. These include a bimodal distribution of stellar flux dispersion and flux variability fraction trends which are highly dependent on stellar magnitude. However, while \citet{Ciardi11} estimated that $\sim$80\% of their G dwarf sample was not significantly variable, this study only finds $\sim$35\% of the starting sample as such. The reason for this disagreement is currently not known.
    \item TESS light curve data is suitable to carry out this type of study, as it was done already for past missions \citep[e.g.][]{Mullally22}. However, TESS has some limitations, including a systematic error floor at 60$ppm$ over 1 hour and a stellar noise strongly increasing as stars get dimmer.
    \item A list of 430 promising stellar calibrators has been selected, which could potentially be used as calibrator targets for the Ariel mission. This method works well in identifying bad calibrator candidates, however, the selected stars are not necessarily stable within the desired flux range. So, these targets should be monitored further to ensure they are still stable close to the launch date, to study short-lived variation features like CMEs and flares, and to better detail their variability level with a higher precision. This can be done by analysing multiple TESS Sector observations, and particularly by using data from observatories with higher precision, like the upcoming ESA mission PLATO.
\end{itemize}


\begin{acknowledgements}
This paper includes data collected by the TESS mission, which are publicly available from the Mikulski Archive for Space Telescopes (MAST). Funding for the TESS mission is provided by NASA’s Science Mission Directorate.\\

This research has also made use of the NASA Exoplanet Archive, which is operated by the California Institute of Technology, under contract with the National Aeronautics and Space Administration under the Exoplanet Exploration Program.\\

Moreover, this research made use of Lightkurve, a Python package for Kepler and TESS data analysis.\\

The authors want to thank Dr. Daphne Stam and Dr. Matthew Kenworthy for their support and supervision during this project. Dr. David Ciardi is thanked for constructive discussions and invaluable help. Finally, we thank Dr. Susan Mullally for the insights on her work.
\end{acknowledgements}

%
%
\bibliographystyle{aa} 
\bibliography{ref} 

\onecolumn
\begin{appendix} 
\section{Light curves and Lomb-Scargle periodograms of sample stars}

\begin{figure}[h]
\centering
    \includegraphics[width=17.5cm]{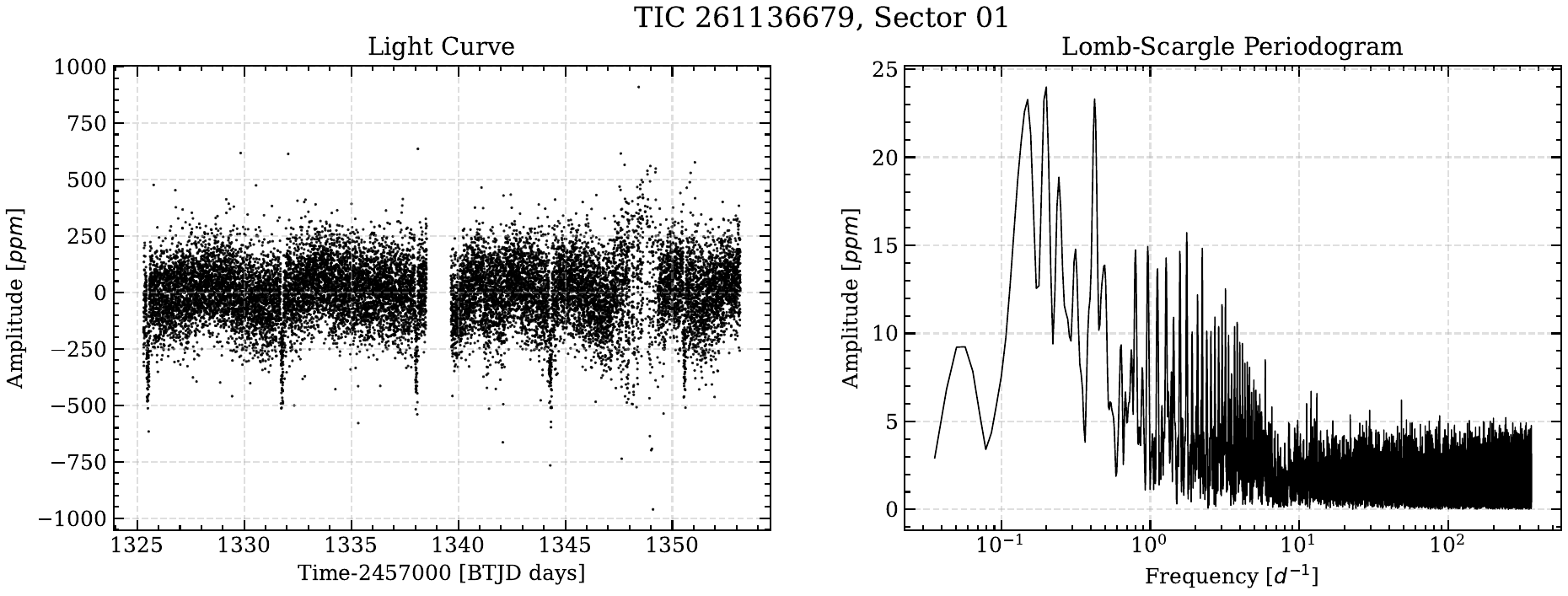}
    \includegraphics[width=17.5cm]{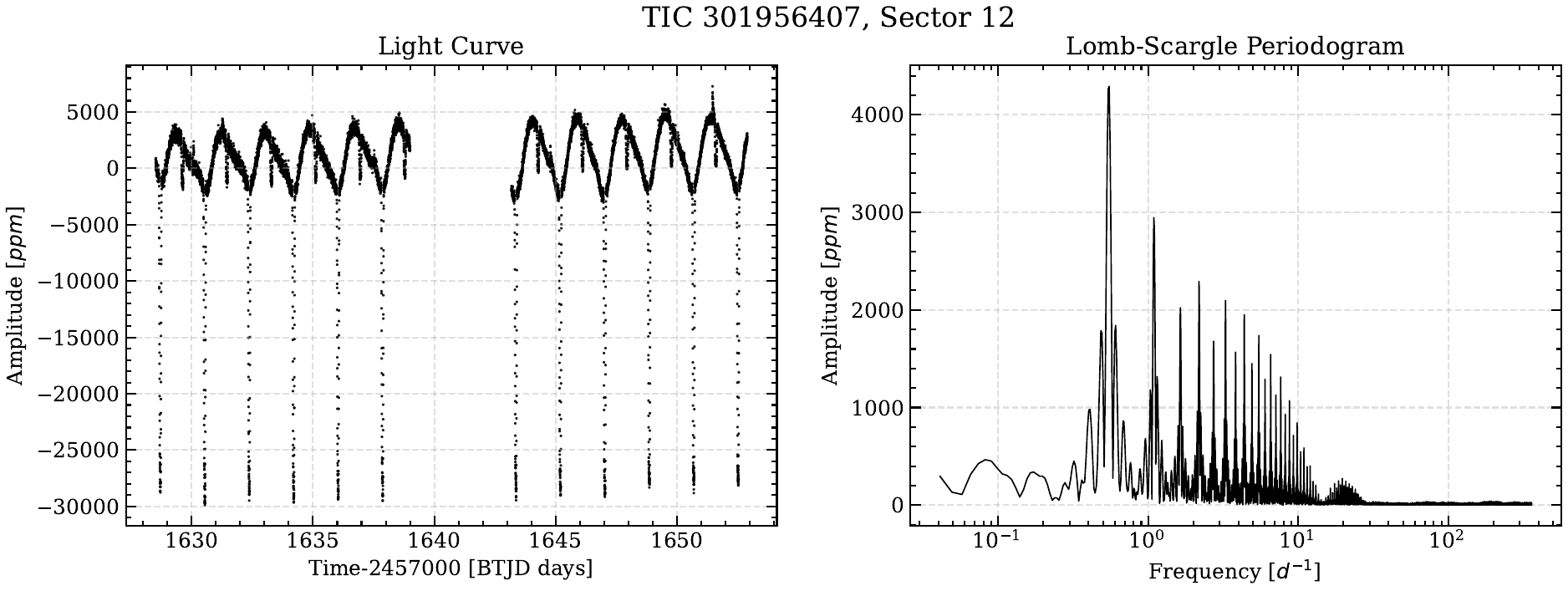}
    \includegraphics[width=17.5cm]{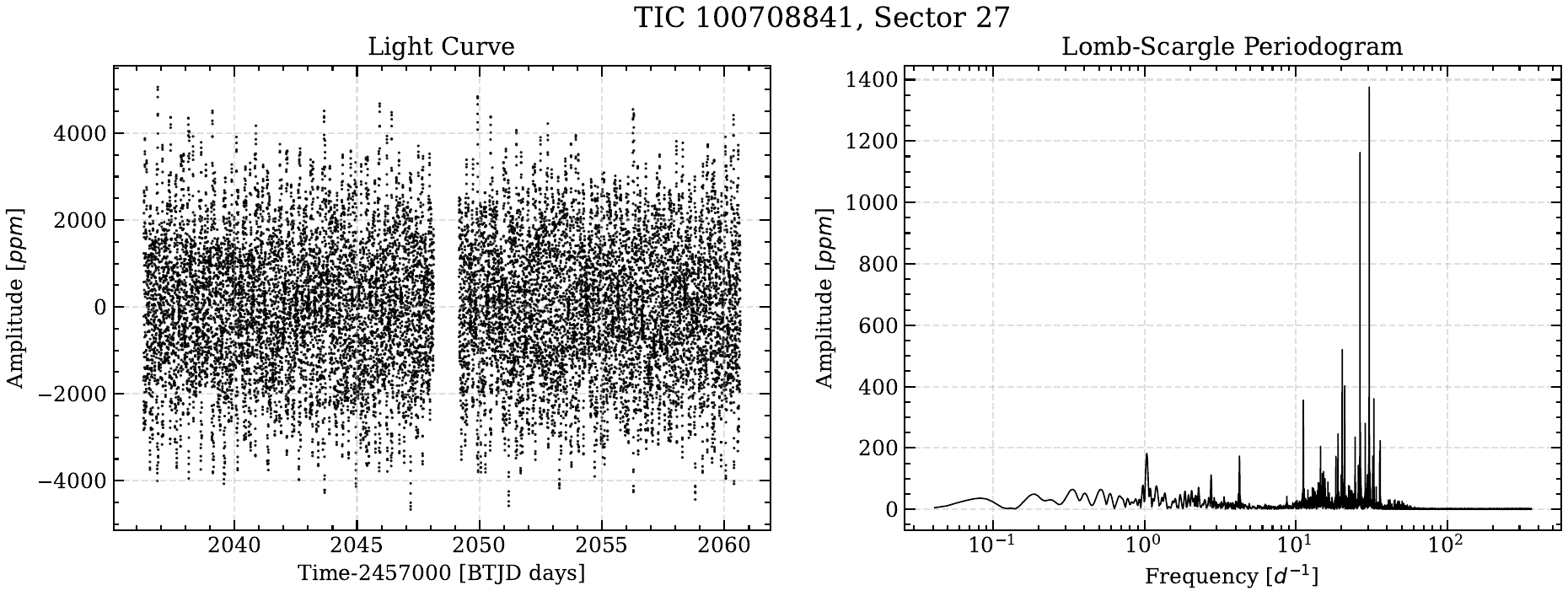}
    \caption{Example light curves and associated LSP of stellar categories: Transit events (exoplanetary transit and eclipsing binaries, upper and middle panel, respectively), and pulsating stars (lower panel).}
    \label{fig:cat_ex_1}
\end{figure}

\newpage

\begin{figure}[h]
\centering
    \includegraphics[width=17.5cm]{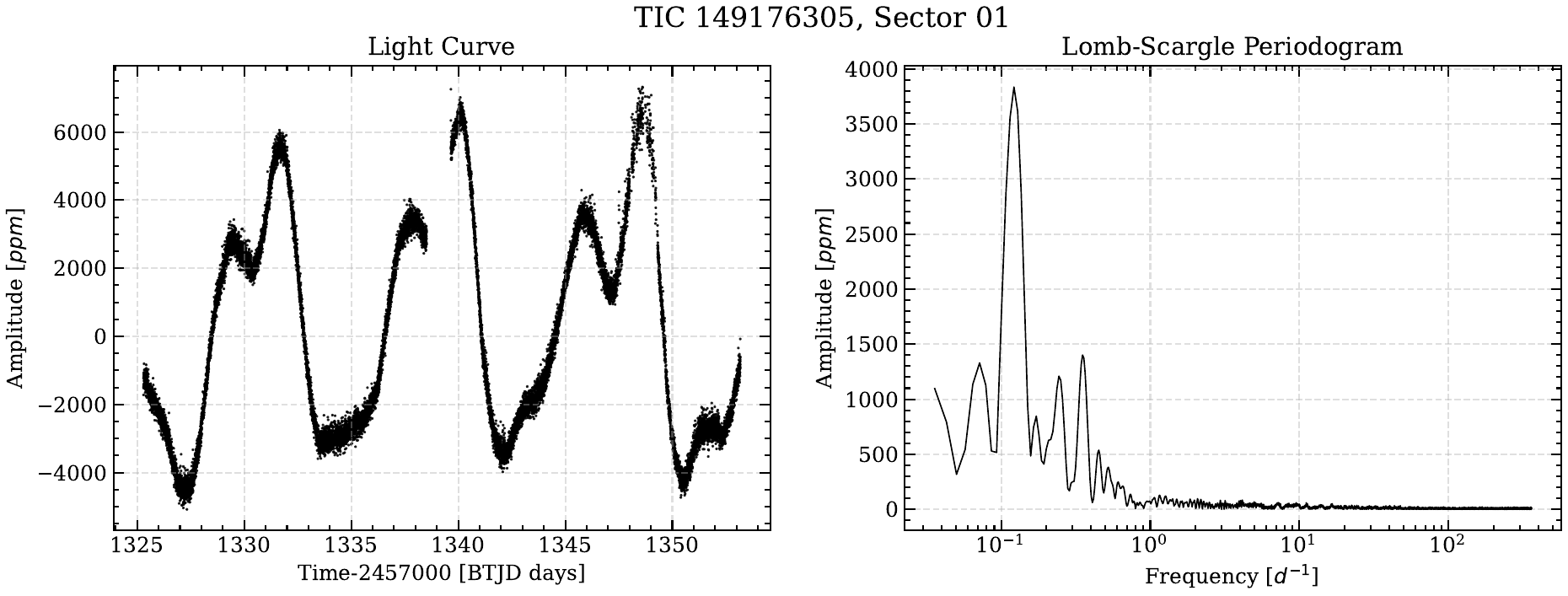}
    \includegraphics[width=17.5cm]{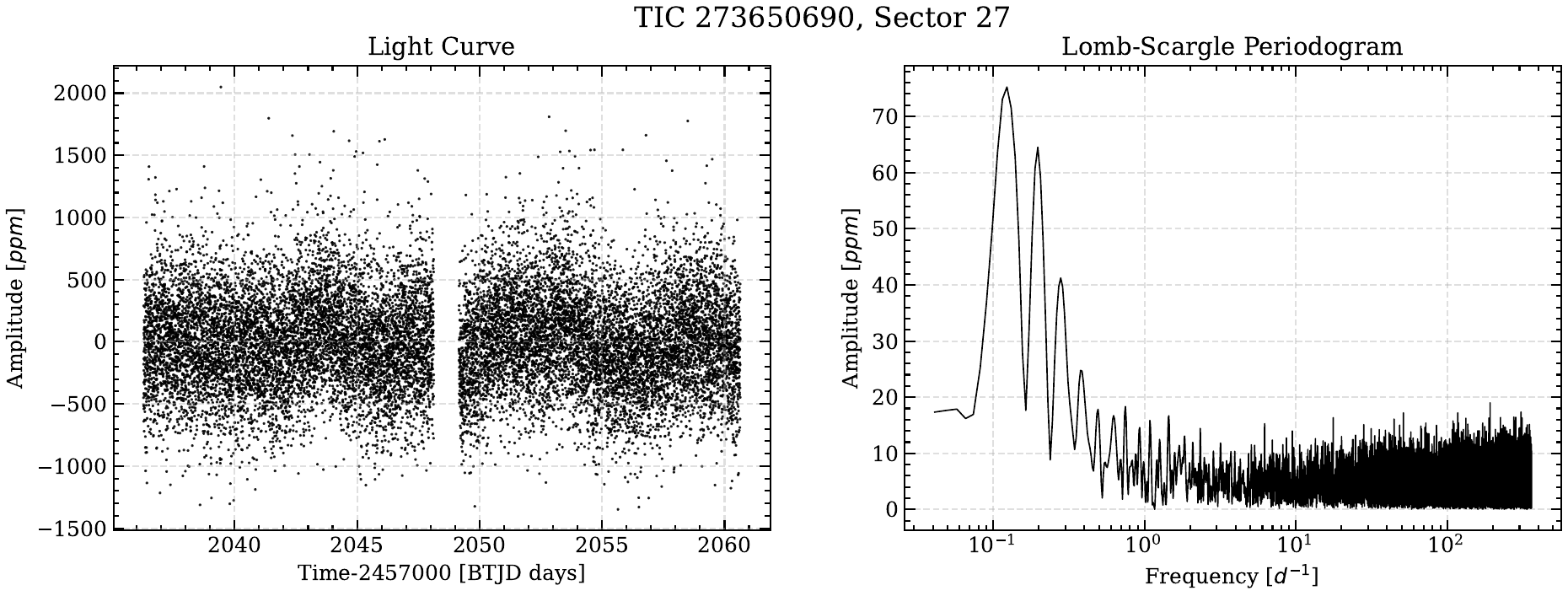}
    \includegraphics[width=17.5cm]{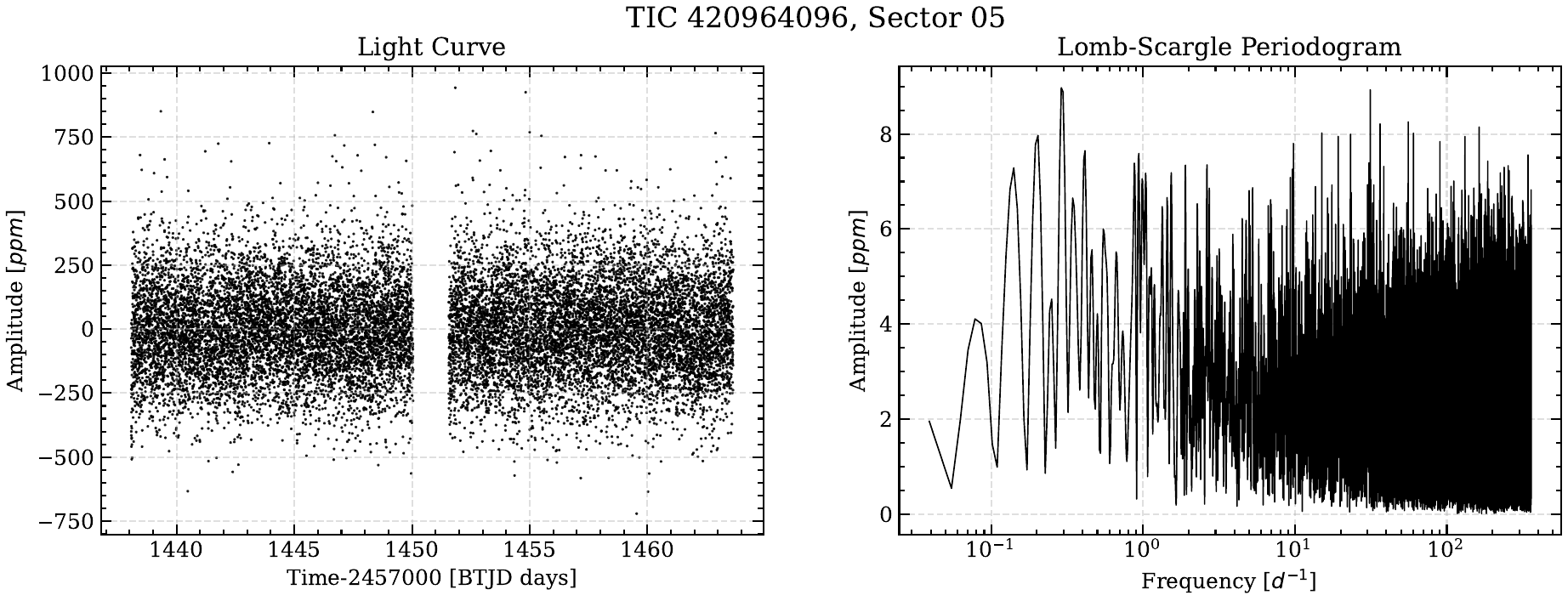}
    \caption{Example light curves and associated LSP of stellar categories: High-amplitude rotating stars (upper panel), low-amplitude rotating stars (middle panel), and low variability (lower panel).}
     \label{fig:cat_ex_2}
\end{figure}

\newpage

\section{Selected stellar calibrators list}
\begin{table}[!h]
\centering
\caption{First twenty objects from the selected stellar calibrator list. Full list available at \url{https://github.com/ElenaTonucci/stellar-calibrators}.}
\label{tab:ariel_calibrators}
\begin{tabular}{lccc}
\hline
TIC identifier & RA & Dec & $\sigma_{Mdn}$ [ppm] \\ \hline
\hline
345176811 & 281.955 & -77.868 & 174.698\\
260823248 & 348.535 & -70.058 & 173.015\\
259886730 &	40.861 & -66.714 & 152.349\\
38469670 &	61.84 &	-64.222 & 171.218\\
76922247 &	271.177 & -59.21 & 206.973\\
220476607 &	75.571 & -56.081 & 213.983\\
155896956 &	4.225 &	-52.651 & 159.913\\
229159785 &	29.25 &	-51.766 & 123.19\\
291635915 &	111.856 & -51.403 & 168.375\\
289074137 &	224.537 & -48.863 & 162.421\\
153027837 &	48.777 & -45.665 & 164.687\\
147113992 &	321.755 & -44.809 & 223.016\\
126584063 &	314.417 & -44.129 & 162.371\\
70829383 &	152.382 & -35.857 & 145.515\\
92498124 &	314.872 & -33.617 & 235.551\\
34798133 &	306.934 & -30.868 & 173.025\\
157324520 &	176.816 & -30.287 & 173.47\\
269885278 &	310.817 & -29.424 & 196.078\\
172734582 &	101.216 & -27.342 & 157.818\\
72748794 &	34.744 & -25.946 & 140.272\\
\hline
\end{tabular}
\end{table}

\end{appendix}

\end{document}